\documentclass[preprint,prd,aps,showpacs,showkeys,nofootinbib]{revtex4}
\usepackage{graphicx}
\usepackage{dcolumn}
\usepackage{bm}
\usepackage{ulem}
\usepackage{color}
\usepackage[dvipsnames]{xcolor}
\usepackage{amssymb}
\usepackage{appendix}
\definecolor{light-gray}{gray}{0.78}
\definecolor{mid-gray}{gray}{0.55}
\definecolor{dark-gray}{gray}{0.32}

\begin{document}

\title{Pair production of $h$ in the $U(1)_X$SSM}
\author{Yue-Tong Liu$^{1,2,3}$, Shu-Min Zhao$^{1,2,3}$\footnote{zhaosm@mail.nankai.edu.cn}, Meng-Zi Cao$^{1,2,3}$, Shuang Di$^{1,2,3}$, Rong-Zhi Sun$^{1,2,3}$, Xing-Xing Dong$^{1,2,3,4}$\footnote{dongxx@hbu.edu.cn}}
\affiliation{$^1$ Department of Physics, Hebei University, Baoding 071002, China}
\affiliation{$^2$ Hebei Key Laboratory of High-precision Computation and Application of Quantum Field Theory, Baoding, 071002, China}
\affiliation{$^3$ Hebei Research Center of the Basic Discipline for Computational Physics, Baoding, 071002, China}
\affiliation{$^4$ Departamento de F\'{\i}sica and CFTP, Instituto Superior T\'{e}cnico, Universidade de Lisboa, Av.Rovisco Pais 1,1049-001 Lisboa, Portugal}
\date{\today}

\begin{abstract}
Higgs pair production provides an important probe of the Higgs self-interaction and the Higgs potential structure. We study the lightest neutral Higgs pair production process $gg \to hh$ via gluon fusion at the 14 TeV LHC, in the $U(1)_X$ supersymmetric standard model. As a $U(1)$ extension of the minimal supersymmetric standard model (MSSM), this model introduces new superfields that bring additional one-loop contributions to the production amplitude. We analyze the parameter dependence of the cross section numerically and present contour distributions in two-dimensional parameter planes. The results indicate that the gauge couplings $g_X$ and $g_{YX}$ are the most sensitive parameters, and the model yields sizable new physics corrections under current experimental constraints. This work helps to understand Higgs physics in the $U(1)_X$SSM and guides new physics searches at the high-luminosity LHC.
\end{abstract}
\maketitle

\section{Introduction}

After the discovery of the Higgs boson\cite{Higg1,NewH1}, extensive theoretical studies have been performed in supersymmetric models\cite{NewH2,NewH3,NewH4,NewH5,NewH6}. One of the main goals has been to measure its properties with the highest possible precision. Current measurements, within experimental and theoretical uncertainties, are consistent with the predictions of the standard model (SM) for the Higgs boson. Higgs pair production provides an intriguing scenario for probing the Higgs self-interaction and searching for new physics. In the SM, the gluon fusion process $gg \to hh$ occurs only at the loop level and its cross section is highly suppressed\cite{LO1,LO2,SMhh1}. Therefore, such new physics scenarios may predict a much larger cross section than the SM result. Experimental searches for Higgs pair production at the LHC are ongoing and provide important constraints on new physics scenarios\cite{EXPhh1}. Early studies of Higgs pair production date back decades, while more recent works extensively investigate this process in various supersymmetric extensions of the SM.

In the MSSM, squark loops and modified Higgs trilinear couplings bring notable corrections to the Higgs pair production cross section\cite{MSSM1,QCD-correction1}. Studies in the NMSSM also show that extra singlet Higgs fields significantly affect this production process\cite{NMSSM1,NMSSM2}. These works confirm that Higgs pair production serves as an effective probe of supersymmetric Higgs sectors. However, the $gg \to hh$ process in the $U(1)_X$SSM still lacks systematic investigation.

MSSM is a widely studied extension of the SM, but it faces several theoretical issues, including the $\mu$-problem\cite{mu1,mu2} and the massless neutrino problem\cite{zwz1,zwz2}. To address these issues, we extend the MSSM with an extra $U(1)_X$ gauge group, with the full gauge symmetry $SU(3)_C \times SU(2)_L \times U(1)_Y \times U(1)_X$\cite{U1x1,U1x2,U1x3,U1x4,GS1}. This model adds three Higgs singlet superfields and right-handed neutrino superfields to the MSSM\cite{Right1,Right2,Higg2}. In the $U(1)_X$SSM, there exist a new gauge boson $A'^X_\mu$ and its supersymmetric partner $\tilde{\lambda}_X$. The model contains five neutral CP-even Higgs component fields, which mix into a $5\times5$ mass-squared matrix and raise the tree-level mass of the lightest CP-even Higgs boson. The additional particles also alleviate the small hierarchy problem in the MSSM. The $\mu$-problem is alleviated by the vacuum expectation value of the singlet field $S$ through the term $\lambda_H \hat{S}\hat{H}_u\hat{H}_d$. In addition, the interaction term $Y_\nu \hat{\nu}\hat{l}\hat{H}_u$ mixes right-handed and left-handed neutrinos, and generates tiny neutrino masses via the seesaw mechanism\cite{TREE}.

In this paper, we investigate the lightest neutral Higgs pair production via gluon fusion $gg \to hh$ at the 14 TeV LHC in the $U(1)_X$SSM. We calculate the one-loop partonic amplitudes for this process, and take all leading-order Feynman diagrams into account. Among these diagrams, triangle diagrams carry the information of trilinear couplings between CP-even Higgs states. We use dimensional regularization to treat the loop integrals. The sum of all leading-order diagrams is ultraviolet finite, as expected for a loop-induced process. We also consider QCD radiative corrections to the cross section, and decompose the next-to-leading order result into leading-order, virtual correction and real-emission correction terms\cite{zsm1,QCD-correction1}. We perform a detailed numerical analysis of the parameter dependence of the total cross section, and present contour plots in two-dimensional parameter planes to show the sensitivity of different parameters clearly.

The outline of this paper is as follows. Section II introduces the basic framework of the $U(1)_X$SSM, including its superpotential and soft supersymmetry-breaking terms. Section III provides the analytical expressions for the scattering amplitude and total cross section of the $gg \to hh$ process. Section IV shows the numerical results and discusses the impacts of key parameters. Section V gives a summary of this work.
\section{The relevant content of $U(1)_X$SSM}

The $U(1)_X$ supersymmetric standard model ($U(1)_X$SSM) is a $U(1)_X$ extension of the minimal supersymmetric standard model. It's gauge group is $SU(3)_C\times SU(2)_L \times U(1)_Y \times U(1)_X$, which retains the full gauge structure of the standard model and the MSSM while introducing one additional $U(1)_X$ gauge factor\cite{U1x1,U1x2,U1x3}. The model is free of gauge anomalies, and its complete superfield content and charge assignments are shown in the Table I.
\begin{table}
\caption{ The superfields in $U(1)_X$SSM}
\begin{tabular}{|c|c|c|c|c|c|c|c|c|c|c|c|}
\hline
Superfields & $\hspace{0.1cm}\hat{q}_i\hspace{0.1cm}$ & $\hat{u}^c_i$ & $\hspace{0.2cm}\hat{d}^c_i\hspace{0.2cm}$ & $\hat{l}_i$ & $\hspace{0.2cm}\hat{e}^c_i\hspace{0.2cm}$ & $\hat{\nu}_i$ & $\hspace{0.1cm}\hat{H}_u\hspace{0.1cm}$ & $\hat{H}_d$ & $\hspace{0.2cm}\hat{\eta}\hspace{0.2cm}$ & $\hspace{0.2cm}\hat{\bar{\eta}}\hspace{0.2cm}$ & $\hspace{0.2cm}\hat{S}\hspace{0.2cm}$ \\
\hline
$SU(3)_C$ & 3 & $\bar{3}$ & $\bar{3}$ & 1 & 1 & 1 & 1 & 1 & 1 & 1 & 1  \\
\hline
$SU(2)_L$ & 2 & 1 & 1 & 2 & 1 & 1 & 2 & 2 & 1 & 1 & 1  \\
\hline
$U(1)_Y$ & 1/6 & -2/3 & 1/3 & -1/2 & 1 & 0 & 1/2 & -1/2 & 0 & 0 & 0  \\
\hline
$U(1)_X$ & 0 & -1/2 & 1/2 & 0 & 1/2 & -1/2 & 1/2 & -1/2 & -1 & 1 & 0  \\
\hline
\end{tabular}
\label{JJ1}
\end{table}

Compared with the MSSM, the $U(1)_X$SSM introduces extra superfields: right-handed neutrino superfields $\hat{\nu}_i$, and three singlet Higgs superfields $\hat{\eta}$, $\hat{\bar{\eta}}$ and $\hat{S}$. Via the tree-level seesaw mechanism, light neutrinos obtain extremely small masses in this framework: the vacuum expectation value of $\hat{\bar{\eta}}$ generates Majorana masses for right-handed neutrinos through the $Y_X \hat{\nu}\hat{\bar{\eta}}\hat{\nu}$ interaction, while the $Y_\nu \hat{\nu}\hat{l}\hat{H}_u$ Yukawa term induces mixing between left-handed and right-handed neutrino states. In the scalar sector, the neutral CP-even components of $H_u$, $H_d$, $\eta$, $\bar{\eta}$ and $S$ mix with each other and form a $5\times 5$ mass-squared matrix.

The superpotential of the $U(1)_X$SSM reads\cite{U1x4,GS2,zsm1,Right1}
\begin{eqnarray}
&&W=l_W\hat{S}+\mu\hat{H}_u\hat{H}_d+M_S\hat{S}\hat{S}-Y_d\hat{d}\hat{q}\hat{H}_d-Y_e\hat{e}\hat{l}\hat{H}_d+\lambda_H\hat{S}\hat{H}_u\hat{H}_d\nonumber
\\&&~~ ~ +\lambda_C\hat{S}\hat{\eta}\hat{\bar{\eta}}+\frac{\kappa}{3}\hat{S}\hat{S}\hat{S}+Y_u\hat{u}\hat{q}\hat{H}_u+Y_X\hat{\nu}\hat{\bar{\eta}}\hat{\nu}
+Y_\nu\hat{\nu}\hat{l}\hat{H}_u.
\end{eqnarray}

The VEVs of the Higgs superfields $H_u$, $H_d$, $\eta$, $\bar{\eta}$ and S are denoted as $v_u$, $v_d$, $v_\eta$, $v_{\bar{\eta}}$ and $v_S$ respectively, and two characteristic mixing angles are defined as $\tan\beta = v_u/v_d$ and $\tan\beta_\eta = v_{\bar{\eta}}/v_\eta$. The vacuum expectation values of the two Higgs doublets and three Higgs singlets are given in component form as follows :
\begin{eqnarray}
&&\hspace{1cm}H_{u}=\left(\begin{array}{c}H_{u}^+\\{1\over\sqrt{2}}\Big(v_{u}+H_{u}^0+iP_{u}^0\Big)\end{array}\right),~~
H_{d}=\left(\begin{array}{c}{1\over\sqrt{2}}\Big(v_{d}+H_{d}^0+iP_{d}^0\Big)\\H_{d}^-\end{array}\right),
\nonumber\\&&\eta={1\over\sqrt{2}}\Big(v_{\eta}+\phi_{\eta}^0+iP_{\eta}^0\Big),~~~
\bar{\eta}={1\over\sqrt{2}}\Big(v_{\bar{\eta}}+\phi_{\bar{\eta}}^0+iP_{\bar{\eta}}^0\Big),~~
S={1\over\sqrt{2}}\Big(v_{S}+\phi_{S}^0+iP_{S}^0\Big).
\end{eqnarray}
where $\phi^0$ and $P^0$ represent the CP-even and CP-odd scalar components respectively.
The soft supersymmetry breaking Lagrangian of the $U(1)_X$SSM contains all soft-breaking terms of the MSSM plus additional terms associated with the new singlet superfields, and it takes the form
\begin{eqnarray}
&&\mathcal{L}_{soft}=\mathcal{L}_{soft}^{MSSM}-B_SS^2-L_SS-\frac{T_\kappa}{3}S^3-T_{\lambda_C}S\eta\bar{\eta}
+\epsilon_{ij}T_{\lambda_H}SH_d^iH_u^j\nonumber\\&&\hspace{1cm}
-T_X^{IJ}\bar{\eta}\tilde{\nu}_R^{*I}\tilde{\nu}_R^{*J}
+\epsilon_{ij}T^{IJ}_{\nu}H_u^i\tilde{\nu}_R^{I*}\tilde{l}_j^J
-m_{\eta}^2|\eta|^2-m_{\bar{\eta}}^2|\bar{\eta}|^2-m_S^2S^2\nonumber\\&&\hspace{1cm}
-(m_{\tilde{\nu}_R}^2)^{IJ}\tilde{\nu}_R^{I*}\tilde{\nu}_R^{J}
-\frac{1}{2}\Big(M_S\lambda^2_{\tilde{X}}+2M_{BB^\prime}\lambda_{\tilde{B}}\lambda_{\tilde{X}}\Big)+h.c.
\end{eqnarray}
where $\mathcal{L}_{soft}^{MSSM}$ stands for the soft SUSY breaking Lagrangian of the MSSM, and h.c. denotes the Hermitian conjugation.

The coexistence of the two Abelian gauge groups $U(1)_Y$ and $U(1)_X$ gives rise to a novel effect absent in the MSSM: gauge kinetic mixing. Even if this mixing vanishes at the grand unification scale $M_{GUT}$, it can be generated through renormalization group evolution. Let $Y^Y$ and $Y^X$ denote the $U(1)_Y$ charge and $U(1)_X$ charge of a superfield respectively. In the original gauge basis, the covariant derivative is written as\cite{U1x5,U1x6,U1x7,U1x11}
\begin{eqnarray}
&&D_\mu=\partial_\mu-i\left(\begin{array}{cc}Y^Y,&Y^X\end{array}\right)
\left(\begin{array}{cc}g_{Y},&g{'}_{{YX}}\\g{'}_{{XY}},&g{'}_{{X}}\end{array}\right)
\left(\begin{array}{c}A{'}_{\mu}^{Y} \\ A{'}_{\mu}^{X}\end{array}\right)\;,
\end{eqnarray}
where $A'^Y_\mu$ and $A'^X_\mu$ represent the gauge fields of $U(1)_Y$ and $U(1)_X$ in the original basis. To simplify the structure of the gauge kinetic term, we introduce a rotation matrix R to perform a basis transformation\cite{U1x5}:
\begin{eqnarray}
&&\left(\begin{array}{cc}g_{Y},&g{'}_{{YX}}\\g{'}_{{XY}},&g{'}_{{X}}\end{array}\right)
R^T=\left(\begin{array}{cc}g_{1},&g_{{YX}}\\0,&g_{{X}}\end{array}\right)~,~~~~
R\left(\begin{array}{c}A_{\mu}^{\prime Y} \\ A_{\mu}^{\prime X}\end{array}\right)
=\left(\begin{array}{c}A_{\mu}^{Y} \\ A_{\mu}^{X}\end{array}\right)\;,
\end{eqnarray}
After the basis rotation, the covariant derivative takes the simplified form
\begin{eqnarray}
&&D_\mu=\partial_\mu-i\left(\begin{array}{cc}Y^Y,&Y^X\end{array}\right)
\left(\begin{array}{cc}g_{1},&g_{{YX}}\\0,&g_{{X}}\end{array}\right)
\left(\begin{array}{c}A_{\mu}^{Y} \\ A_{\mu}^{X}\end{array}\right)\;.
\end{eqnarray}

Here $g_X$ is the gauge coupling constant of the $U(1)_X$ group, and $g_{YX}$ characterizes the kinetic mixing strength between the $U(1)_Y$ and $U(1)_X$ sectors.

At tree level, three neutral gauge bosons: the $U(1)_Y$ gauge boson $A^Y_\mu$, the $SU(2)_L$ neutral gauge boson $V^3_\mu$, and the $U(1)_X$ gauge boson $A^X_\mu$  mix with one another\cite{U1x3}.

The explicit tree-level mass matrices and Feynman rules for the $U(1)_X$SSM, which follow directly from the superpotential and soft SUSY-breaking terms, are collected in appendix A.

\section{Scattering Amplitudes and Cross Sections}

\begin{figure}[ht]
\setlength{\unitlength}{3mm}
\centering
\includegraphics[width=2in]{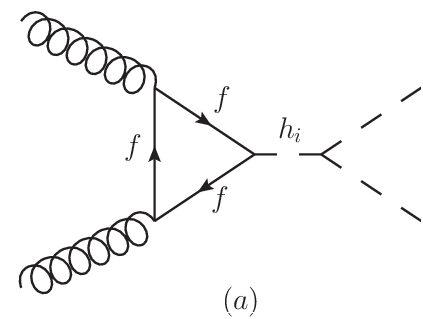}
\setlength{\unitlength}{5mm}
\centering
\includegraphics[width=2in]{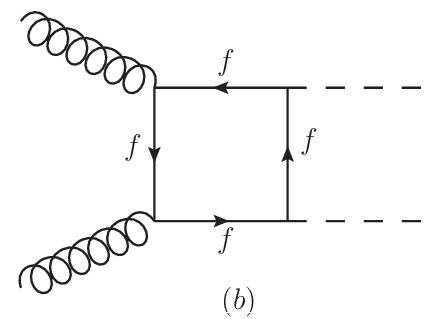}
\setlength{\unitlength}{3mm}
\centering
\includegraphics[width=2in]{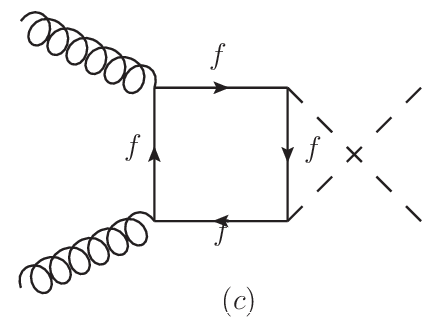}
\setlength{\unitlength}{3mm}
\centering
\includegraphics[width=2in]{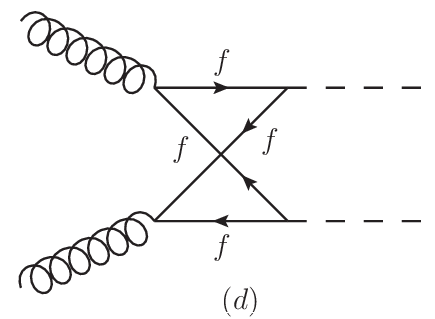}
\setlength{\unitlength}{3mm}
\centering
\includegraphics[width=2in]{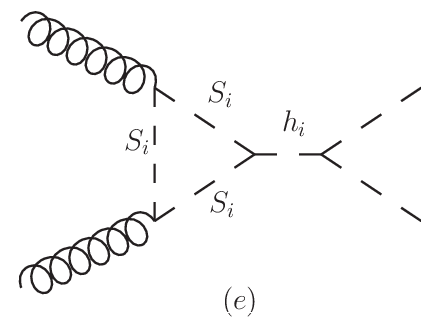}
\setlength{\unitlength}{3mm}
\centering
\includegraphics[width=2in]{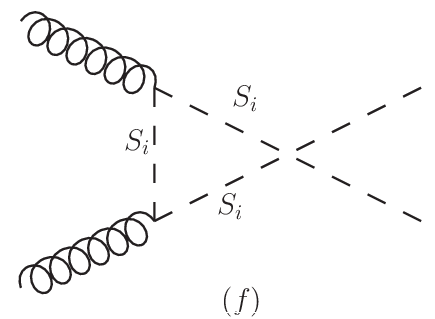}
\setlength{\unitlength}{3mm}
\centering
\includegraphics[width=2in]{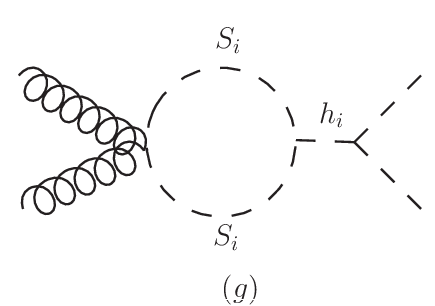}
\setlength{\unitlength}{3mm}
\centering
\includegraphics[width=2in]{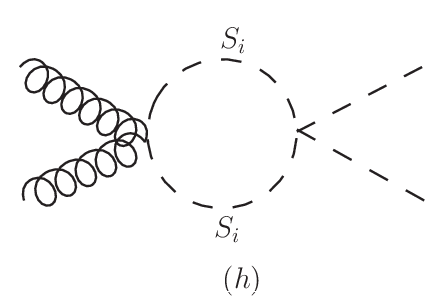}
\setlength{\unitlength}{3mm}
\centering
\includegraphics[width=2in]{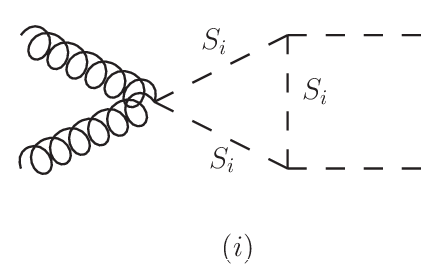}
\setlength{\unitlength}{3mm}
\centering
\includegraphics[width=2in]{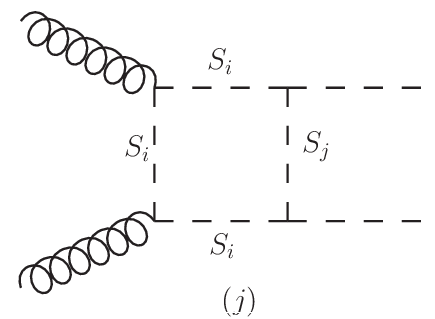}
\setlength{\unitlength}{3mm}
\centering
\includegraphics[width=2in]{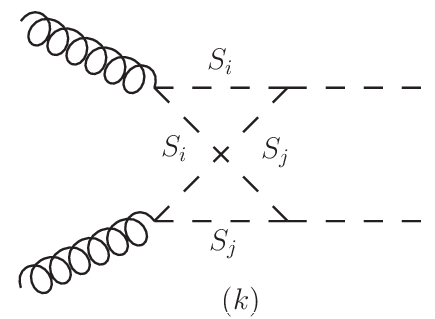}
\setlength{\unitlength}{3mm}
\centering
\includegraphics[width=2in]{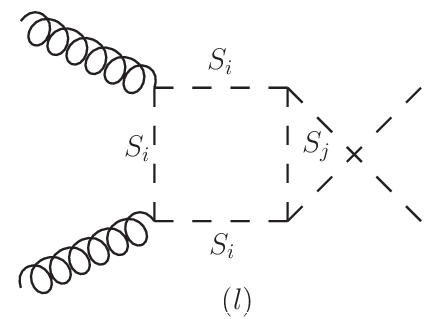}
\caption{Feynman diagrams for the $gg \rightarrow hh$ process in the $U(1)_X$SSM. $h_i$ are CP-even Higgs, $f$ are top and bottom quarks and $S_{i(j)}$ are stop and sbottom quarks.}
\label{num1}
\end{figure}

We investigate the pair production cross section of the lightest Higgs boson through the gluon fusion process $gg \to hh$ at $\sqrt{s} = 14$ TeV, using one-loop amplitudes for the partonic process $g(p_1)g(p_2) \to h(p_3)h(p_4)$ in supersymmetric extensions of the SM. All leading-order Feynman diagrams that contribute to the $gg \to hh$ channel are presented in Fig.\ref{num1}, among which triangle diagrams encode the trilinear couplings between CP-even Higgs states. With the top and bottom Yukawa couplings determined by separate experimental measurements, the pair production cross section data can be used to explore the correlation between the next-to-lightest neutral Higgs mass $m_{h_2}$ and the trilinear Higgs couplings $C_{hhh}$ and $C_{h_2hh}$. Here $h_2$ denotes the next-to-lightest neutral Higgs eigenstate, \(C_{hhh}\) stands for the triple self-coupling of the lightest neutral Higgs, and $C_{h_2hh}$ represents the coupling of one next-to-lightest neutral Higgs boson to two lightest neutral Higgs bosons.

For the calculation of the polarized cross section, we introduce the explicit polarization vectors for gluons with helicities ${(\lambda_1,\;\lambda_2)}$:
\begin{eqnarray}
&&\epsilon^\mu_1(p_1,\lambda_1=\pm1)=\frac{1}{\sqrt{2}}(0,\mp1,-i,0)\;,\nonumber\\
&&\epsilon^\mu_2(p_2,\lambda_2=\pm1)=\frac{1}{\sqrt{2}}(0,\pm1,-i,0)\;.
\label{polarization vectors}
\end{eqnarray}

In the center-of-mass frame, the four-momenta of initial and final state particles are parameterized by the beam energy E and scattering angle $\theta$:
\begin{eqnarray}
&&p^\mu_1=E(1,0,0,-1)\;,\nonumber\\
&&p^\mu_2=E(1,0,0,1)\;,\nonumber\\
&&p^\mu_3=E(1,-\sqrt{1-\frac{4m^2_h}{\hat{s}}} \sin\theta,0,-\sqrt{1-\frac{4m^2_h}{\hat{s}}} \cos\theta)\;,\nonumber\\
&&p^\mu_4=E(1,\sqrt{1-\frac{4m^2_h}{\hat{s}}} \sin\theta,0,\sqrt{1-\frac{4m^2_h}{\hat{s}}} \cos\theta)\;.
\label{momenta}
\end{eqnarray}
where $p_{1,2}$ are the incoming momenta of the two colliding gluons, and $p_{3,4}$ are the outgoing momenta of the two produced lightest Higgs bosons.

The Mandelstam variables for this $2\to2$ scattering process are defined as follows:
\begin{eqnarray}
\hat{s}=(p_1+p_2)^2=(p_3+p_4)^2\;,\nonumber\\
\hat{t}=(p_1-p_3)^2=(p_2-p_4)^2\;,\nonumber\\
\hat{u}=(p_1-p_4)^2=(p_2-p_3)^2\;.
\label{Mandelstam variables}
\end{eqnarray}

The leading-order cross section for the partonic subprocess $gg \to hh$ is expressed as an integral over the squared helicity amplitudes in phase space:
\begin{eqnarray}
&&{\hat{\sigma}}=\int^{\hat t_{max}}_{\hat t_{min}}\; d\hat{t} \;\frac{1}{4096\pi \hat{s}^2}
\;{(|\sum_{n}M^{(n)}_{++}|^2+|\sum_{n}M^{(n)}_{+-}|^2
+|\sum_{n}M^{(n)}_{--}|^2+|\sum_{n}M^{(n)}_{-+}|^2)}\;,\nonumber\\
\label{subprocess cross section}
\end{eqnarray}
with the upper and lower integration boundaries given by
\begin{eqnarray}
{\hat t_{min}}=(m^2_h-\frac{\hat{s}}{2})-\frac{1}{2}\;\sqrt{1-\frac{4m^2_h}{\hat{s}}}\;\hat{s}\;,\nonumber\\
{\hat t_{max}}=(m^2_h-\frac{\hat{s}}{2})+\frac{1}{2}\;\sqrt{1-\frac{4m^2_h}{\hat{s}}}\;\hat{s}\;.
\end{eqnarray}

Here $M_{\lambda_1\lambda_2}^{(n)}$ is the helicity amplitude for the lightest neutral Higgs pair production from the n-th Feynman diagram. CP conservation imposes the relations $M_{++}=M_{--}$ and $M_{+-}=M_{-+}$, where the $\pm$ subscripts label the two independent helicity states of the initial gluons.
The differential cross section for the hadronic process $gg \to hh$ is obtained by convolving the partonic cross section with the gluon-gluon parton luminosity:
\begin{eqnarray}
\frac{d\sigma}{d\sqrt{\hat{s}}}=\frac{2\sqrt{\hat{s}}}{s}\:{\hat\sigma(gg-hh)}\:\frac{dL_{gg}}{d\tau}\;,
\label{differential cross section}
\end{eqnarray}
where $\tau = \hat{s}/s$, and the gluon-gluon parton luminosity takes the form
\begin{eqnarray}
\frac{dL_{gg}}{d\tau}=\int^1_\tau\;\frac{dx}{x}\;f_g(x,\mu_{_F})\;f_g(\frac{\tau}{x},\mu_{_F})\;.
\label{luminosity}
\end{eqnarray}

In this expression, $f_g(x, \mu_F)$ is the gluon parton distribution function, $x$ is the momentum fraction carried by the gluon inside the proton, and $\mu_F$ is the factorization scale.
The total leading-order cross section for lightest neutral Higgs pair production via gluon fusion in proton-proton collisions reads
\begin{eqnarray}
\sigma_{LO}(pp\rightarrow gg\rightarrow hh)=\int^1_{\tau_0}\:d\tau
\:{\hat\sigma(gg\rightarrow hh)}\:\frac{dL_{gg}}{d\tau}\;,
\label{cross section}
\end{eqnarray}
with the lower integration limit $\tau_0 = (2m_h)^2/s$.
QCD radiative corrections are known to produce considerable corrections to the theoretical prediction for the lightest Higgs pair production cross section. The next-to-leading order QCD corrections\cite{SMhh1} to gluon-fusion Higgs pair production are computed in the heavy-top limit\cite{SMhh1}. The full NLO cross section can be decomposed into the LO contribution plus virtual and real correction terms:
\begin{eqnarray}
\sigma_{NLO}=\sigma_{LO}+\Delta\sigma_{virt}+\Delta\sigma_{gg}+\Delta\sigma_{gq}+\Delta\sigma_{q\bar{q}}\;,
\label{NLO cross section}
\end{eqnarray}
where the LO cross section is given by Eq.(\ref{cross section}). The virtual and real-emission correction terms take the form:
\begin{eqnarray}
&&\Delta\sigma_{virt}=\frac{(\alpha_s{(\mu_{_R})})}{\pi}\:\int^1_{\tau_0}\:d\tau\:
\frac{dL_{gg}}{d\tau}\:{\hat\sigma}_{LO}(\hat{s}=\tau{s})\:C_{virt}(\hat{s})\:,
\nonumber\\
&&\Delta\sigma_{ij}=\frac{(\alpha_s{(\mu_{_R})})}{\pi}\:\int^1_{\tau_0}\:d\tau\:
\frac{dL_{ij}}{d\tau}\:\int^1_{\frac{\tau_0}{\tau}}\:\frac{dz}{z}\:
{\hat\sigma}_{LO}(\hat{s}=z\tau{s})\:C_{ij}(\hat{s},z)\:.
\label{QCD correction}
\end{eqnarray}

The functions $C_{ij}(\hat{s}, z)$ (with $ij = gg,\,gq,\,q\bar{q}$) take the form\cite{SMhh1}
\begin{eqnarray}
&&C_{gg}(\hat{s},z)=-zP_{gg}(z)\:\log\frac{\mu^2_{_F}}{\tau s}
\nonumber\\
&&\hspace{2.2cm}
+6[1+z^4+(1-z)^4]\:{(\frac{\log(1-z)}{1-z})_+}\:+d_{gg}(z)\:,
\nonumber\\
&&C_{gq}(\hat{s},z)=-\frac{z}{2}P_{gq}(z)\:\log\frac{\mu^2_{_F}}{\tau s(1-z)^2}+d_{gq}(z)\:,
\nonumber\\
&&C_{q\bar{q}}(\hat{s},z)=d_{q\bar{q}}(z)\:.
\label{QCD correction1}
\end{eqnarray}
The $d_{ij}(z)$ terms represent the finite hard parts of the real-emission corrections, and they contain no collinear divergences. Their explicit forms are given as
\begin{eqnarray}
d_{gg}(z)=-\frac{11}{2}(1-z)^3,
\nonumber\\
d_{gq}(z)=\frac{2}{3}z^2-(1-z)^2,
\nonumber\\
d_{q\bar{q}}(z)=\frac{32}{27}(1-z)^3.
\end{eqnarray}
Here $P_{ij}(z)$ ($i,j = g,q,\bar{q}$) are the Altarelli--Parisi splitting functions\cite{QCD}.
\begin{eqnarray}
P_{gg}(z) &=& 6\left\{ \left(\frac{1}{1-z}\right)_+
+\frac{1}{z}-2+z(1-z) \right\} + \frac{33-2N_F}{6}\delta(1-z),
\nonumber \\
P_{gq}(z) &=& \frac{4 (1+(1-z)^2)}{3 z}.
\end{eqnarray}

\section{Numerical analysis}

In this section, we investigate the amplitude and cross section of \(gg\to hh\). The corresponding Feynman diagrams are depicted in Fig.\ref{num1}. We take one diagram in Fig.\ref{num1} as an example for illustration. The Feynman amplitude for Fig.\ref{num1}(e) reads:
\begin{eqnarray}
&&\mathcal{M}_{(e)}=\int\frac{d^Dk}{(2\pi)^D}\frac{1}{[(p_1+p_2)^2-m_{{h}_i}^2][(p_1+p_2+k)^2-m_{\tilde{u}_i}^2][(p_2+k)^2-m_{\tilde{u}_i}^2](k^2-m_{\tilde{u}_i}^2)} \nonumber\\&&\hspace{1.5cm}A_{HHH}A_{H \tilde{U} \tilde{U}}g_3^2(T^aT^b+T^bT^a)g^{\alpha\beta}\delta^{ij}\epsilon(p_1)_\alpha^a\epsilon(p_2)_\beta^b.\label{x0}
\end{eqnarray}

In this expression, $p_1$ and $p_2$ denote the momenta of the two incoming gluons, and k is the loop momentum. $m_{h_i}$ corresponds to the mass of the i-th CP-even Higgs boson, and $m_{\tilde{u}_i}$ is the mass of the up-type squark. $\mathcal{L}_{HHH}$ stands for the vertex factor of the Higgs trilinear coupling, while $\mathcal{L}_{H\tilde{U}\tilde{U}}$ represents the vertex factor for the coupling of a Higgs boson to a pair of up-type squarks. $g_3$ is the strong coupling constant, and $\epsilon(p_1)^a_\alpha$ and $\epsilon(p_2)^b_\beta$ are the polarization vectors of the two incoming gluons, respectively.

We begin with the evaluation of the Feynman integral. Following the parametrization formula for three denominators given in Ref.\cite{Du4}, we express the product of three propagators in the form
\begin{eqnarray}
\frac{1}{ABC} = \int_{0}^{1}dx\int_{0}^{1}2ydy\frac{1}{[(Ax + B(1 - x))y + C(1 - y)]^3}.\label{x1}
\end{eqnarray}

This integration procedure substantially enhances the efficiency of numerical computations in our work. We arrive at the expression
\begin{eqnarray}
&&\int_{0}^{1}dx\int_{0}^{1}2ydy\Big\{\Big[(k^2-m_{\tilde{u}_i}^2)x + [(p_2+k)^2-m_{\tilde{u}_i}^2](1 - x)\Big] y\nonumber\\&&\hspace{1.8cm}
 + [(p_1+p_2+k)^2-m_{\tilde{u}_i}^2](1 - y)\Big\}^{-3}\nonumber\\&&\hspace{1.8cm}
=\int_{0}^{1}dx\int_{0}^{1}2ydy\frac{1}{(k'^2 - R^2)^3}.
\end{eqnarray}

The auxiliary quantities involved in the momentum shift are defined as
\begin{eqnarray}
&&p_1 + p_2 - p_1y - p_2xy = T,\nonumber\\&&
k + T = k',\nonumber\\&&
R^2 = m_{\tilde{u}_i}^2(-1 + y) - m_{\tilde{u}_i}^2y + 2p_1p_2(-1 + y)(-2 + xy).
\end{eqnarray}

We then evaluate the momentum integral in D-dimensional space using the standard formula for the denominator structure:
\begin{eqnarray}
&&\int\frac{d^Dk^\prime}{(2\pi)^D}\frac{(k^2)^\alpha}{(k^2-R^2)^\beta}=i\frac{(-1)^{\alpha-\beta}}{(4\pi)^{\frac{D}{2}}}\frac{\Gamma(1+\frac{D}{2})\Gamma(\beta-\alpha-\frac{D}{2})}
{\Gamma(\frac{D}{2})\Gamma(\beta)(R^2)^{\beta-\alpha-\frac{D}{2}}}.
\end{eqnarray}

Dimensional regularization is applied to treat the loop integrals, where the spacetime dimension takes the value $d = 4 - 2\epsilon$ and the limit $d \to 4$ is taken. The sum of all diagrams is ultraviolet finite, so no renormalization is required at this order.

We take the following experimental limits into account in our numerical study:
\begin{enumerate}
    \item The mass of the lightest CP-even Higgs boson $m_{h}$ agrees with the experimental result $m_{h} = 125.13 \pm 0.11\ \text{GeV}$\cite{pdg}.
    \item To match LHC experimental data, the value of $\tan\beta_\eta$ must be less than 1.5\cite{tanb1.5}.
    \item According to the latest LHC data\cite{LHC1,LHC2}, and both up-squarks and down-squarks are heavier than 1500 GeV.
    \item The $Z'$ boson mass $M_{Z'}$ is larger than 5.1 TeV, and the ratio $M_{Z'}/g_B$ satisfies $M_{Z'}/g_X \geq 6\ \text{TeV}$ \cite{U1x3}.
\end{enumerate}

All parameter values used in our numerical calculation meet all the experimental limits listed above.
\subsection{one-dimensional line graph}

We use plots to visualize the impact of variables on the results, with the quantitative parameters set as follows
\begin{eqnarray}
&&T_{\lambda_H}=0.6\;{\rm TeV},~ T_{\lambda_C}=-0.1\;{\rm TeV},~ v_S=4.3\;\rm{TeV},~ \tan\beta=20,\nonumber\\
&&v_\eta=17\times\cos\theta_\eta~{\rm TeV},
~ v_{\bar\eta}=17\times\sin\theta_\eta~{\rm TeV},~
l_W=5\;{\rm TeV^2},\nonumber\\
&&B_\mu=B_S=5\;{\rm TeV^2},~T_\kappa=3\;{\rm TeV},~ Y_{X11} = Y_{X22} = Y_{X33} = 1, \nonumber\\
&& m_{\tilde{Q}11}^2 = m_{\tilde{Q}22}^2 = 1.9\;{\rm TeV^2},~ m_{\tilde{Q}33}^2 =6\;{\rm TeV^2},~m_{\tilde{U}11}^2 = m_{\tilde{U}22}^2 = 2.5\;{\rm TeV^2},\nonumber\\
&&m_{\tilde{U}33}^2 =6\;{\rm TeV^2},~
m_{\tilde{D}11}^2 = m_{\tilde{D}22}^2 = m_{\tilde{D}33}^2 = 2.7\;{\rm TeV^2}, \nonumber\\
&&\tan\beta_\eta = \frac{v_{\bar\eta}}{v_\eta} = 1.1,~ T_{u11} = T_{u22} = 1\;{\rm TeV},~ T_{u33} = 4.25\;{\rm TeV},\nonumber\\
&&T_{d11} = T_{d22} = T_{d33} = 0.2\;{\rm TeV},~ \alpha_s(M_Z) = 0.118.
\label{cs}
\end{eqnarray}

\begin{figure}
\setlength{\unitlength}{5mm}
\centering
\includegraphics[width=3in]{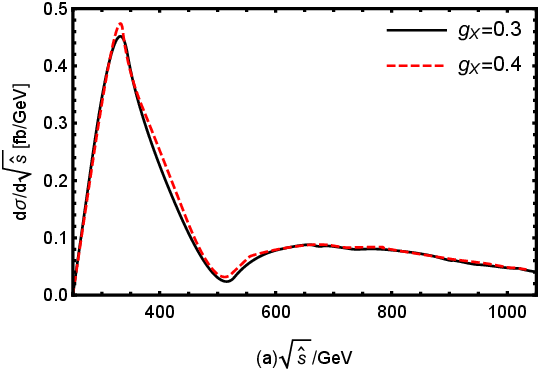}
\setlength{\unitlength}{5mm}
\centering
\includegraphics[width=3.1in]{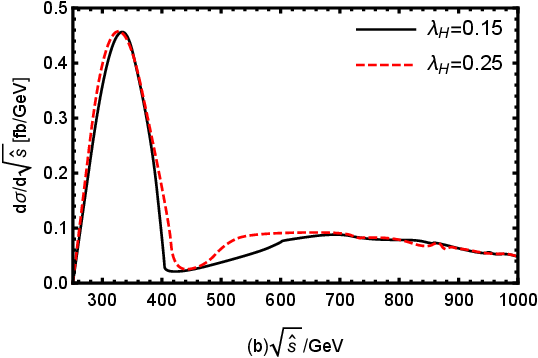}
\setlength{\unitlength}{5mm}
\centering
\includegraphics[width=3.1in]{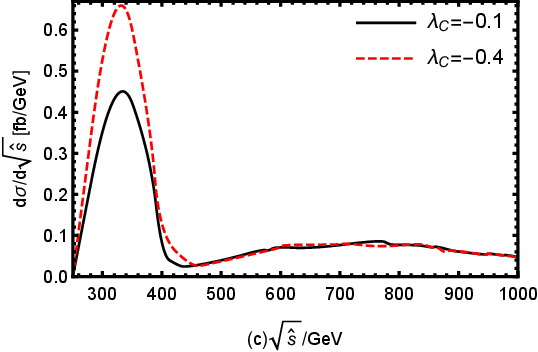}
\caption{The total differential cross section as a function of the partonic center-of-mass energy \(\sqrt{\hat{s}}\) in the \(U(1)_X\) model.}
\label{num2}
\end{figure}

In the $U(1)_X$SSM with the parameters $g_{YX}=0.15$, $\mu=800\;{\rm{GeV}}$, $\kappa=0.1$ and $M_S=3600\;{\rm{GeV}}$, we present the total differential cross section as a function of the partonic center-of-mass energy $\sqrt{\hat{s}}$ in Fig.\ref{num2}(a)-2(c), which illustrate the impacts of $g_X$, $\lambda_H$ and $\lambda_C$ on the cross section respectively.
All curves show typical resonant behavior, with a sharp peak in the low-energy region, a clear dip structure after the peak, and a slowly falling flat tail at high energies.

We note that varying a single parameter also modifies the full Higgs mass spectrum and mixing angles, so we plot relevant diagrams for the numerical results.
Fig.\ref{num2}(a) shows the differential cross section for different values of $g_X$, with $\lambda_H = 0.1$ and $\lambda_C = -0.1$ fixed.
The solid line corresponds to $g_X = 0.3$ and the dashed line corresponds to $g_X = 0.4$.
As the gauge coupling of the $U(1)_X$ group, $g_X$ directly enters the loop through vertices and mass matrices of new physics particles, and makes important contribution to the loop amplitude.
As $g_X$ increases, the differential cross section decreases uniformly over the energy range $200\ \text{GeV} < \sqrt{\hat{s}} < 500\ \text{GeV}$.
When $\sqrt{\hat{s}}$ exceeds 500 GeV, the impact of $g_X$ becomes insignificant and the cross section approaches a nearly constant behavior.

Fig.\ref{num2}(b) presents the dependence of the differential cross section on $\lambda_H$ with $\lambda_C = -0.1$ and $g_X = 0.3$ fixed.
The solid line stands for $\lambda_H = 0.15$ and the dashed line stands for $\lambda_H = 0.25$.
$\lambda_H$ controls the mixing between the singlet Higgs fields and the two Higgs doublets, and affects the cross section mainly through the changes in Higgs matrix and couplings.
Since it acts as a subleading effect in this parameter region, the differential cross section shows only a moderate increase as $\lambda_H$ increases.
The influence of $\lambda_H$ becomes insignificant when $\sqrt{\hat{s}}$ exceeds 600 GeV.

Fig.\ref{num2}(c) displays the variation of the differential cross section with $\lambda_C$ for fixed $g_X = 0.3$ and $\lambda_H = 0.1$.
The two curves correspond to $\lambda_C = -0.1$ and $\lambda_C = -0.4$ respectively.
The coupling $\lambda_C$ directly modifies the trilinear interactions among the three singlet Higgs fields, which dominates the s-channel resonant triangle diagrams.
In the energy range $0 < \sqrt{\hat{s}} < 450\ \text{GeV}$, the resonant peak rises dramatically and becomes slightly narrower as the absolute value of $\lambda_C$ increases.
The impact of $\lambda_C$ becomes insignificant when $\sqrt{\hat{s}}$ exceeds 450 GeV.

\begin{figure}
\setlength{\unitlength}{5mm}
\centering
\includegraphics[width=3in]{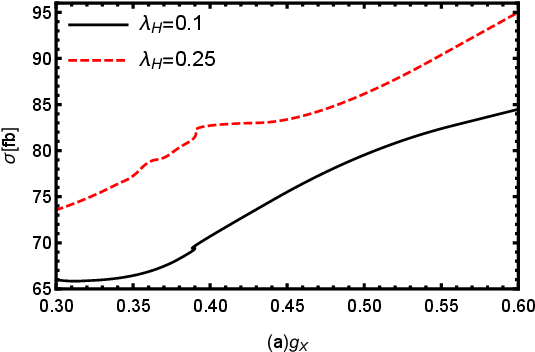}
\setlength{\unitlength}{5mm}
\centering
\includegraphics[width=3in]{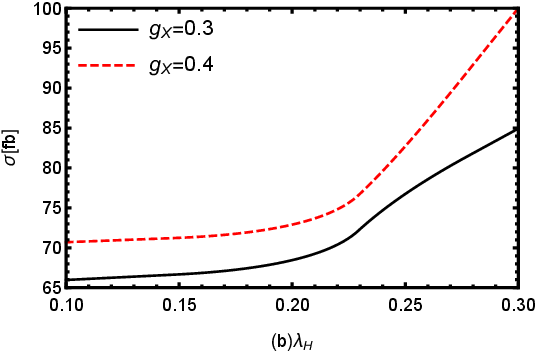}
\setlength{\unitlength}{5mm}
\centering
\includegraphics[width=3in]{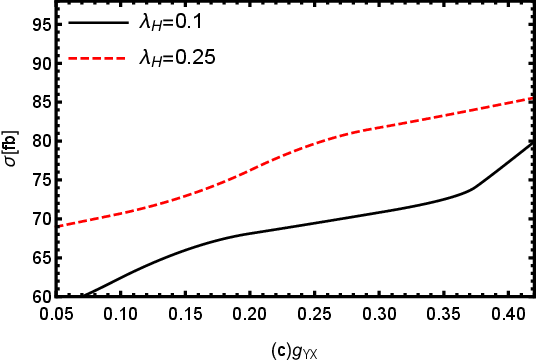}
\setlength{\unitlength}{5mm}
\centering
\includegraphics[width=3in]{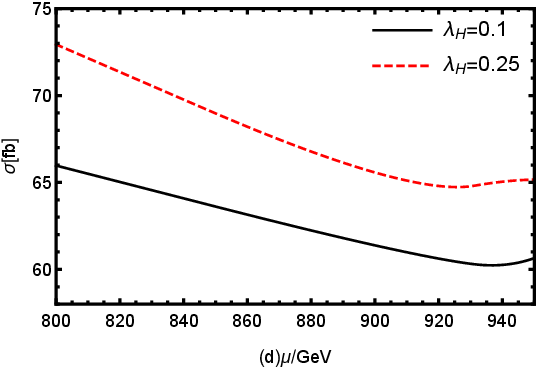}
\setlength{\unitlength}{5mm}
\centering
\includegraphics[width=3in]{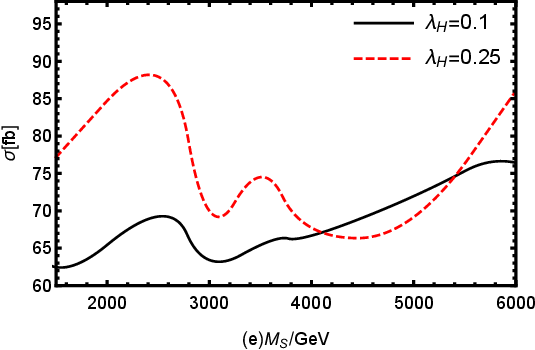}
\caption{Dependence of the total Higgs pair production cross section on the key parameters $g_X$, $\lambda_H$, $g_{YX}$, $\mu$ and $M_S$ in the \(U(1)_X\)SSM, with Fig.\ref{num3}(a)-(e) illustrating the variation corresponding to each parameter respectively.}
\label{num3}
\end{figure}

We further examine the dependence of the total production cross section on the key parameters of the $U(1)_X$ model, as shown in Fig.\ref{num3}(a)-(e). Each panel shows how the cross section changes with one model parameter, with all other baseline parameters fixed at $\kappa=0.1$ and $\lambda_C = -0.1$. In each plot, the solid and dashed lines represent two different values of this parameter. Fig.\ref{num3}(a)-(e) show how the total production cross section depends respectively on the key parameters $g_X$, $\lambda_H$, $g_{YX}$, $\mu$ and $M_S$ in the $U(1)_X$SSM.
In Fig.\ref{num3}(b), which displays the cross section versus $\lambda_H$, the solid line stands for $g_X=0.3$ and the dashed line stands for $g_X=0.4$. For Fig.\ref{num3}(c), (d) and (e), where the cross sections are plotted against $g_{YX}$, $\mu$ and $M_S$ respectively, the black solid and red dashed lines in each panel represent $\lambda_H=0.1$ and $\lambda_H=0.25$.

$g_X$ is the gauge coupling constant of the new $U(1)_X$ gauge group. It enters the mass matrices of multiple particles (neutralino, down-squark, up-squark, neutral Higgs, charged Higgs) and several coupling vertices
($h\bar{\chi}_i^0\chi_j^0$ $h\tilde{u}^*_i\tilde{u}_j$ $h\tilde{d}^*_i\tilde{d}_j$)and it can enhance new physics effects. Obviously, $g_X$ is a sensitive parameter. Taking the reference parameters $g_{YX}=0.15$, $\mu=800\;{\rm{GeV}}$ and $M_S=3600\;{\rm{GeV}}$ we display the total cross section as a function of $g_X$ in Fig.\ref{num3}(a), with the black solid and red dashed lines corresponding to $\lambda_H=0.1$ and $\lambda_H=0.25$. We find that for $g_X$ in the range of 0.3 to 0.6, the values of both lines grow as $g_X$ increases.

$\lambda_H$ comes from the $\lambda_H S H_u H_d$ term in the superpotential. It appears in the mass matrices of many particles (chargino, neutralino, down-squark, up-squark, neutral Higgs, charged Higgs), and it may bring complex effects to the numerical results. The contribution of each diagram in Fig.\ref{num3}(b) is affected by $\lambda_H$. Similarly, for $g_{YX}=0.15$, $\mu=800\;{\rm{GeV}}$ and $M_S=3600\;{\rm{GeV}}$, the dependence of the total cross section on the coupling  $\lambda_H$ is presented in Fig.\ref{num3}(b). For $\lambda_H$ from 0.1 to 0.3, the total cross section rises as $\lambda_H$ increases. When $\lambda_H > 0.23$, the total cross section increases significantly with the increase of $\lambda_H$.

In Fig.\ref{num3}(c), we plot the total cross section versus $g_{YX}$ with fixed $g_X=0.3$, $\mu=800\;{\rm{GeV}}$, $M_S=3600\;{\rm{GeV}}$, and the total cross section increases steeply with the rise of $g_{YX}$. In the $U(1)_X$SSM, $g_{YX}$ is a mixed gauge coupling beyond the MSSM. We change $g_{YX}$ from 0.05 to 0.42, and the cross section also becomes larger as $g_{YX}$ increases. The behavior in Fig.\ref{num3}(a) is similar to that in Fig.\ref{num3}(c), which means $g_X$ and $g_{YX}$ have comparable effects to some extent.

The $\mu$ parameter is the higgsino mass term in the superpotential, which enters the mass matrices of charginos, neutralinos and sfermions, and affects the loop amplitudes by modifying the particle mass spectra and mixing angles. In Fig.\ref{num3}(d), with $g_X = 0.3$, $g_{YX} = 0.15$ and $M_S=3600\;{\rm{GeV}}$, we illustrate the variation of the total production cross section with $\mu$ for two different $\lambda_H$ values. For both curves, the total cross section decreases as $\mu$ increases and then levels off gradually in the high-$\mu$ range. This behavior can be understood from the fact that $\mu$ enters the sfermion mass matrices through the left-right mixing terms. As $\mu$ increases, the sfermion masses are shifted and their contributions to the loop diagrams are modified, leading to the observed decrease of the cross section. At large $\mu$, the cross section becomes insensitive to further increases since the dominant squark contributions are already decoupled. The curve corresponding to $\lambda_H = 0.25$ lies consistently above the $\lambda_H = 0.1$ curve, since a larger $\lambda_H$ results in a higher total cross section, following the same trend as observed in Fig.\ref{num3}(a) and Fig.\ref{num3}(c).

$M_S$ is the characteristic mass scale of the $U(1)_X$ singlet Higgs sector, which directly determines the masses of singlet-like Higgs bosons and their mixing with the doublet components, thus playing a crucial role in the s-channel resonant production. In Fig.3(e), we display the total cross section as a function of $M_S$ by keeping $g_X = 0.3$, $\lambda_H = 0.1$, $g_{YX} = 0.15$ and $\mu=800\;{\rm{GeV}}$ unchanged. The cross section curves show rich resonant structures: a sharp main resonance peak appears around $2500\;{\rm{GeV}}$, followed by a weaker secondary peak near $3500\;{\rm{GeV}}$; after a dip region, the cross section exhibits a steady rising trend at large $M_S$. These multiple resonance features originate from the propagation of different singlet Higgs eigenstates in the s-channel triangle diagrams. The slow rise at high $M_S$ results from the competition between the propagator mass suppression and the coupling modification induced by the singlet-doublet mixing.
\subsection{filled contour plot}

\begin{figure}
\setlength{\unitlength}{5mm}
\centering
\includegraphics[width=3in]{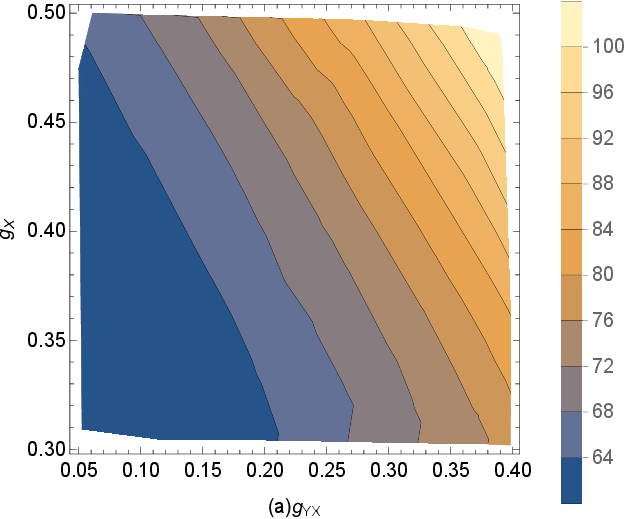}
\setlength{\unitlength}{5mm}
\centering
\includegraphics[width=3.1in]{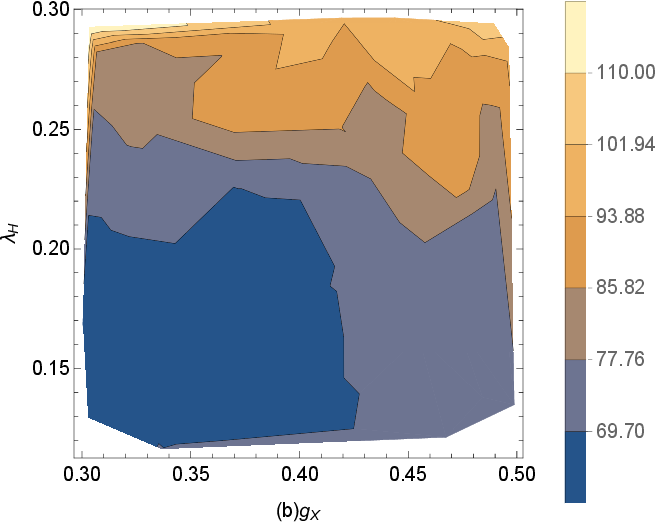}
\setlength{\unitlength}{5mm}
\centering
\includegraphics[width=3.1in]{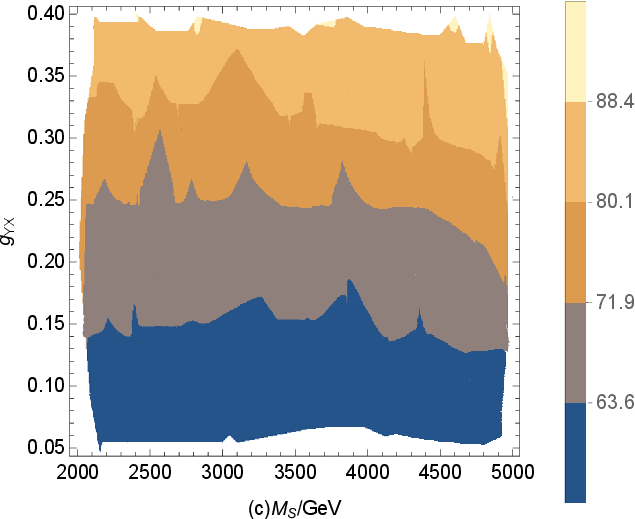}
\setlength{\unitlength}{5mm}
\centering
\includegraphics[width=3in]{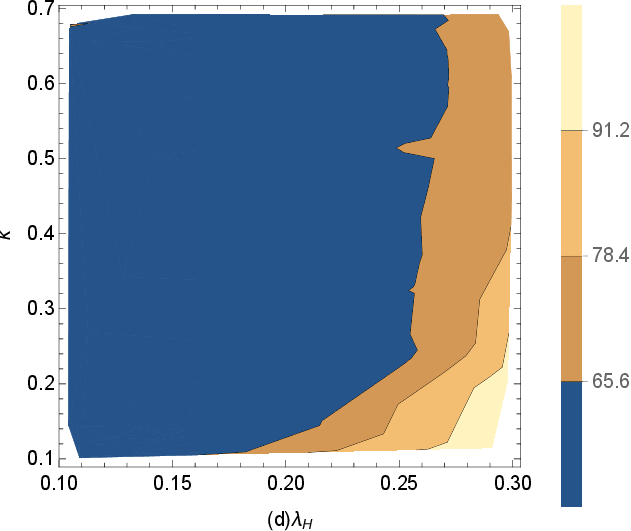}
\setlength{\unitlength}{5mm}
\centering
\includegraphics[width=3in]{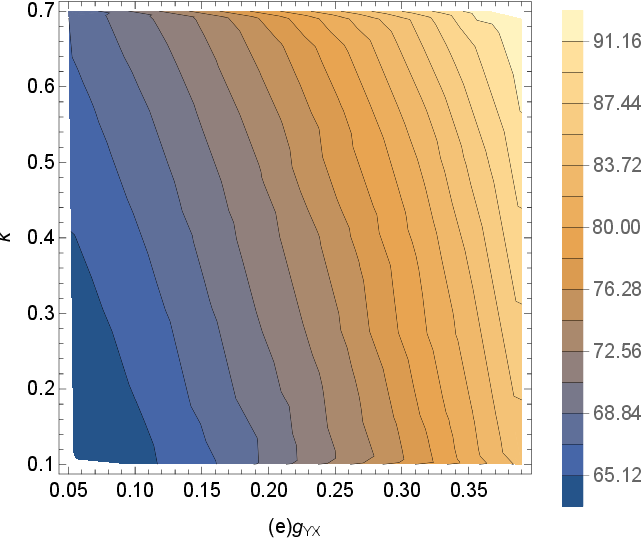}
\caption{Contour plots of the total cross section for the $gg \to hh$ process in the $U(1)_X$SSM. Fig.\ref{num4}(a)-(e) display the cross section distributions in five different two-parameter planes respectively, and the color bars on the right give the corresponding cross section values.}
\label{num4}
\end{figure}

To gain a clear picture of how model parameters affect the $gg \to hh$ production process, we present the contour plots of the total cross section in Fig.\ref{num4}. Each plot shows the cross section distribution in the plane of two free parameters for fixed $\mu=800\;{\rm{GeV}}$ and $\lambda_C = -0.1$, with the color scale on the right marking the magnitude of the cross section.

In Fig.\ref{num4}(a), with $\lambda_H = 0.1$, $\kappa=0.1$ and $M_S=3600\;{\rm{GeV}}$ fixed, the horizontal axis spans $0.05 \le g_{YX} \le 0.40$ and the vertical axis spans $0.30 \le g_X \le 0.50$. The total cross section rises clearly with the increase of both $g_{YX}$ and $g_X$, with the minimum value in the lower-left corner and the maximum value in the upper-right corner of the plot. Physically, both $g_{YX}$ and $g_X$ are gauge couplings that directly enter the interaction vertices and mass matrices of new physics particles. Both $g_X$ and $g_{YX}$ are gauge couplings unique to the $U(1)_X$ model, and the total cross section rises as these two couplings increase and falls as they decrease. They modify the loop amplitudes of the $gg \to hh$ process, so both parameters show strong and comparable sensitivity to the cross section.

Fig.\ref{num4}(b) presents the cross section in the $g_X$--$\lambda_H$ plane, for fixed $g_{YX} = 0.15$, $\kappa=0.1$ and $M_S=3600\;{\rm{GeV}}$, with $0.30 \le g_X \le 0.50$ on the horizontal axis and $0.15 \le \lambda_H \le 0.30$ on the vertical axis. The total cross section grows with increasing $g_X$ and $\lambda_H$ over the scanned parameter range. This is because $g_X$ directly adjusts the coupling strength of supersymmetric particles running in the loops. By contrast, $\lambda_H$ affects the cross section mainly through Higgs mixing, which is a subleading effect in this parameter region.

In Fig.\ref{num4}(c), with $g_X=0.3$, $\lambda_H = 0.1$ and $\kappa=0.1$ kept unchanged, we choose $M_S$ as the horizontal axis in the range of $2000 \le M_S \le 5000\ \text{GeV}$ and $g_{YX}$ as the vertical axis in the range of $0.05 \le g_{YX} \le 0.40$. The cross section increases significantly with larger $g_{YX}$, but it changes only slightly when $M_S$ varies across the whole range, showing a weak dependence on $M_S$. This feature can be explained as follows: $g_{YX}$ directly modifies the strength of the relevant coupling vertices, so it changes the loop amplitude in an obvious way. Each plot adopts a different set of benchmark parameters, so the dependence of the total cross section on the scanned parameters varies accordingly.

Fig.\ref{num4}(d) displays the cross section behavior in the $\lambda_H$--$\kappa$ plane, with all remaining parameters fixed at $g_X=0.3$, $g_{YX} = 0.15$ and $M_S=3600\;{\rm{GeV}}$, where $\lambda_H$ ranges from 0.10 to 0.30 and $\kappa$ ranges from 0.10 to 0.70. The cross section increases steadily with growing $\lambda_H$. The influence of $\kappa$ is mainly visible in the large $\lambda_H$ region, and it barely affects the cross section when $\lambda_H$ is small. $\lambda_H$ plays a dominant role in this parameter plane, because it directly shapes the Higgs potential and modifies Higgs mixing. The effect of $\kappa$ on the total cross section becomes more evident when $\lambda_H$ takes larger values.

In Fig.\ref{num4}(e), for $g_X=0.3$, $\lambda_H = 0.1$ and $M_S=3600\;{\rm{GeV}}$ held constant, the parameter plane is formed by $g_{YX}$ ($0.05 \le g_{YX} \le 0.35$) and $\kappa$ ($0.10 \le \kappa \le 0.70$). Both parameters can enhance the total cross section as their values increase, and the variation trend along the $g_{YX}$ direction is more remarkable than that along the $\kappa$ direction. Again, $g_{YX}$ is the more sensitive parameter than $\kappa$. In this condition, the effect of $\kappa$ is relatively weak, so it leads to a much milder change in the cross section.

As illustrated by the contour distributions in Fig.\ref{num4}, the total cross section falls in the range of 60--100 fb across the scanned parameter space. The parameter dependence trends and the overall cross section magnitude are fully consistent with the one-dimensional results shown in Fig.\ref{num3}, which verifies the robustness of our numerical calculations.

Overall, the gauge couplings $g_X$, $g_{YX}$ and the Higgs coupling $\lambda_H$ are the most sensitive parameters for the $gg \to hh$ cross section in this model. The soft mass $M_S$ and the coupling $\kappa$ have relatively weak impacts on the cross section. Based on the above calculations and contour analyses, the total $gg \to hh$ cross section in the $U(1)_X$SSM can exceed 60 fb within the scanned parameter range.

In the decoupling limit with heavy squarks and heavy singlet fields, our calculation reduces to the SM result, confirming the consistency of our approach.

\section{Conclusion}

In this paper, we study the $gg \to hh$ Higgs pair production via gluon fusion at the 14 TeV LHC in the $U(1)_X$SSM. We compute the one-loop amplitudes and hadronic cross sections, and analyze the parameter dependence of the differential and total cross sections numerically. The differential cross section shows a typical resonant behavior: a sharp low-energy peak, a clear dip structure, and a slowly falling tail at high energy, and the couplings $g_X$, $\lambda_H$ and $\lambda_C$ can modify the peak height and position significantly. For the total cross section, $g_X$ and $g_{YX}$ are the most sensitive parameters, while $\lambda_H$ and $\mu$ have moderate effects. The gaugino mass $M_S$ and coupling $\kappa$ have weak impacts: they mainly affect the Higgs mass and mixing, which enter the loop amplitudes only through subleading channels. The two-dimensional contour plots confirm these sensitivity patterns clearly.

In the SM, the total cross section for $gg\to hh$ at the 14 TeV LHC lies in the range of 35--40 fb, which is widely accepted as the theoretical benchmark\cite{SM}. In the MSSM, the total cross section is generally enhanced compared with the SM result due to additional contributions from squark loops and heavy Higgs resonance effects, and its value varies considerably across different parameter configurations\cite{MSSM}. In the $U(1)_X$SSM investigated in this work, the total cross section typically ranges from 60 fb to 100 fb within the allowed parameter region, and the specific value depends on parameter choices, and is clearly larger than the SM result.

All parameter sets in our work satisfy current experimental constraints, including the Higgs mass measurement, the lower mass limits of supersymmetric particles and the $Z'$ boson, and other LHC bounds. Within the allowed parameter region, the $U(1)_X$ model introduces significant new physics corrections to the $gg \to hh$ cross section relative to the SM. Our results show that Higgs pair production serves as an effective probe of the $U(1)_X$SSM, and future high-luminosity LHC measurements will further constrain the parameter space and provide insights into new physics beyond the SM.

\appendix
\section{Explicit Tree-Level Mass Matrices and Feynman Rules in the $U(1)_X$SSM}

The symmetric mass-squared matrix of gauge bosons in the basis ($A_{\mu}^{Y}$, $V_{\mu}^{3}$, $A_{\mu}^{X}$) is
\begin{eqnarray}
&&\left(\begin{array}{*{20}{c}}
\frac{1}{8}g_{1}^2 v^2 &~~~ -\frac{1}{8}g_{1}g_{2} v^2 & ~~~\frac{1}{8}g_{1}(g_{YX}+g_X) v^2 \\
-\frac{1}{8}g_{1}g_{2} v^2 &~~~ \frac{1}{8}g_{2}^2 v^2 & ~~~~-\frac{1}{8}g_{2}(g_{YX}+g_X) v^2\\
\frac{1}{8}g_{1}(g_{YX}+g_X) v^2 &~~~ -\frac{1}{8}g_{2}(g_{YX}+g_X) v^2 &~~~~ \frac{1}{8}(g_{YX}+g_X)^2 v^2+\frac{1}{8}g_{{X}}^2 \xi^2
\end{array}\right),\label{gauge matrix}
\end{eqnarray}
with the definitions $v^2 = v_u^2 + v_d^2$ and $\xi^2 = v_\eta^2 + v_{\bar{\eta}}^2$. Diagonalization of this mass matrix yields three mass eigenvalues: one corresponds to the massless photon, and the other two correspond to the physical $Z$ and $Z'$ bosons:
\begin{eqnarray}
&&m_\gamma^2=0,\nonumber\\
&&m_{Z,{Z^{'}}}^2=\frac{1}{8}\Big((g_{1}^2+g_2^2+(g_{YX}+g_X)^2)v^2+4g_{X}^2\xi^2\nonumber\\
&&\mp\sqrt{(g_{1}^2+g_{2}^2+(g_{YX}+g_X)^2)^2v^4+8((g_{YX}+g_X)^2-g_{1}^2-
g_{2}^2)g_{X}^2v^2\xi^2+16g_{X}^4\xi^4}\Big).
\end{eqnarray}

In the basis $\left(\tilde{d}_{L}, \tilde{d}_{R}\right)$, the mass-squared matrix for down-type squarks takes the form
\begin{equation}
M^2_{\tilde{D}} = \left(
\begin{array}{cc}
m_{\tilde{d}_L\tilde{d}_L^{*}} &m^\dagger_{\tilde{d}_R\tilde{d}_L^{*}}\\
m_{\tilde{d}_L\tilde{d}_R^{*}} &m_{\tilde{d}_R\tilde{d}_R^{*}}\end{array}
\right),
\end{equation}
with the explicit expressions of the matrix elements as follows\cite{Right2,Higg2,TREE,zsm1}:
\begin{eqnarray}
&&m_{\tilde{d}_L\tilde{d}_L^{*}} = \frac{1}{24}
 \Big( (3 g_{2}^{2}  + g_{1}^{2} + g_{Y X}^{2}+  g_{Y X} g_{X}) ( v_{u}^{2}- v_{d}^{2}  ) +  2g_{Y X} g_{X}
 ( v_{\bar{\eta}}^{2}  - v_{\eta}^{2})\Big)+m_{\tilde{Q}}^2  +\frac{ v_{d}^{2}}{2} {Y_{d}^2},\nonumber\\
&&m_{\tilde{d}_L\tilde{d}_R^{*}} = -\frac{1}{2}  \Big(\sqrt{2}  (- v_d T_d  + v_u Y_d \mu ) + v_u v_S Y_d {\lambda}_{H} \Big),\nonumber\\
&&m_{\tilde{d}_R\tilde{d}_R^{*}} = \frac{1}{24}   \Big(  (2 g_{1}^{2} + 2g_{Y X}^{2}+ 5g_{Y X} g_{X}+3g_{X}^{2})
 ( v_{u}^{2}  - v_{d}^{2})+ 2(2g_{Y X} g_{X}+3 g_{X}^{2} )( v_{\bar{\eta}}^{2}  - v_{\eta}^{2})\Big)\nonumber\\&&\hspace{1.8cm} +m_{\tilde{D}}^2  +\frac{ v_{d}^{2}}{2} {Y_{d}^2}.
\end{eqnarray}

This matrix is diagonalized by $Z^D$:
\begin{equation}
Z^D M^2_{\tilde{D}} Z^{D,\dagger} = m_{2,\tilde{d}}^{\text{dia}}.
\end{equation}

In the basis $\left(\tilde{u}_{L}, \tilde{u}_{R}\right)$, the mass-squared matrix for up-type squarks is given by
\begin{equation}
M^2_{\tilde{U}} = \left(
\begin{array}{cc}
m_{\tilde{u}_L\tilde{u}_L^{*}} &m^\dagger_{\tilde{u}_R\tilde{u}_L^{*}}\\
m_{\tilde{u}_L\tilde{u}_R^{*}} &m_{\tilde{u}_R\tilde{u}_R^{*}}\end{array}
\right),
\end{equation}
where the matrix elements read
\begin{eqnarray}
	&&m_{\tilde{u}_L\tilde{u}_L^{*}} = \frac{1}{24}  \Big( (g_{1}^{2} -3 g_{2}^{2}+ g_{Y X}^{2}+   g_{Y X} g_{X}) ( v_{u}^{2}- v_{d}^{2}  )
 +   g_{Y X} g_{X}  (2 v_{\bar{\eta}}^{2}  -2 v_{\eta}^{2} )\Big)
+  m_{\tilde{Q}}^2  +\frac{ v_{u}^{2}}{2} {Y_{u}^2},\nonumber\\
	&&m_{\tilde{u}_L\tilde{u}_R^{*}} = -\frac{1}{2}   \Big(\sqrt{2}  (v_d Y_u \mu  - v_u T_u ) + v_d v_S Y_u {\lambda}_{H} \Big),\nonumber\\
	&&m_{\tilde{u}_R\tilde{u}_R^{*}} = \frac{1}{24}   \Big( (4g_{1}^{2} + 4g_{Y X}^{2}+  7g_{Y X} g_{X}+3  g_{X}^{2})
( v_{d}^{2}- v_{u}^{2})+  2(4g_{Y X} g_{X} +3  g_{X}^{2}) (  v_{\eta}^{2}-v_{\bar{\eta}}^{2}  )\Big) \nonumber \\
	&&\hspace{1.8cm}+  m_{\tilde{U}}^2  +\frac{ v_{u}^{2}}{2} {Y_{u}^2}.\nonumber
\end{eqnarray}
This matrix is diagonalized by $Z^U$:
\begin{equation}
Z^U M^2_{\tilde{U}} Z^{U,\dagger} = m_{2,\tilde{u}}^{\text{dia}}.
\end{equation}

At the tree level, the mass-squared matrix for the CP-even Higgs
($\phi_d,\phi_u,\phi_\eta,\bar\phi_\eta,\phi_s$) is as follows
\begin{eqnarray}
&&M^2_h = \left(\begin{array}{ccccc}m_{\phi_d\phi_d}&m_{\phi_u\phi_d}&m_{\phi_\eta\phi_d}
&m_{\phi_{\bar\eta}\phi_d}&m_{\phi_s\phi_d}\\
m_{\phi_d\phi_u}&m_{\phi_u\phi_u}&m_{\phi_\eta\phi_u}
&m_{\phi_{\bar\eta}\phi_u}&m_{\phi_s\phi_u}\\
m_{\phi_d\phi_\eta}&m_{\phi_u\phi_\eta}&m_{\phi_\eta\phi_\eta}
&m_{\phi_{\bar\eta}\phi_\eta}&m_{\phi_s\phi_\eta}\\
m_{\phi_d\phi_{\bar\eta}}&m_{\phi_u\phi_{\bar\eta}}&m_{\phi_\eta\phi_{\bar\eta}}
&m_{\phi_{\bar\eta}\phi_{\bar\eta}}&m_{\phi_s\phi_{\bar\eta}}\\
m_{\phi_d\phi_s}&m_{\phi_u\phi_s}&m_{\phi_\eta\phi_s}
&m_{\phi_{\bar\eta}\phi_s}&m_{\phi_s\phi_s}\end{array}\right),\indent
\end{eqnarray}
\begin{eqnarray}
&&m_{\phi_d\phi_d} = m_{H_d}^2 + \mu^2
+ \frac{1}{8}\Big([g_1^2 + (g_X + g_{YX})^2 + g_2^2](3v_d^2 - v_u^2)\nonumber\\&&\hspace{1.8cm}+2(g_{YX}g_X + g_X^2)(v_\eta^2 - v_{\bar\eta}^2)\Big) + \sqrt{2}v_S\mu\lambda_H + \frac{1}{2}(v_u^2 + v_S^2)\lambda_H^2,\nonumber\\
&&m_{\phi_d\phi_u} = -\frac{1}{4}\Big(g_2^2 + (g_{YX} + g_X)^2 + g_1^2\Big)v_dv_u + \lambda_H^2v_dv_u - \lambda_H l_W\nonumber\\&&\hspace{1.8cm} -\frac{1}{2}\lambda_H v_\eta v_{\bar\eta}\lambda_C + v_S^2\kappa - B_\mu - \sqrt{2}v_S(\frac{1}{2}T_{\lambda_H}+M_S\lambda_H),\nonumber\\
&&m_{\phi_u\phi_u} = m_{H_u}^2 + \mu^2 + \frac{1}{8}\Big\{\Big([g_1^2 + (g_X + g_{YX})^2 + g_2^2]\Big)(3v_u^2 - v_d^2)\nonumber\\&&\hspace{1.8cm}+2(g_{YX}g_X + g_X^2)( v_{\bar\eta}^2 - v_\eta^2 )\Big\} + \sqrt{2}v_S\mu\lambda_H + \frac{1}{2}(v_d^2 + v_S^2)\lambda_H^2,\nonumber\\
&&m_{\phi_d\phi_\eta} = \frac{1}{2}g_X(g_{YX} + g_X)v_dv_\eta -\frac{1}{2}v_uv_{\bar\eta}\lambda_H\lambda_C,\nonumber\\
&&m_{\phi_u\phi_\eta} = -\frac{1}{2}g_X(g_{YX} + g_X)v_dv_\eta -\frac{1}{2}v_dv_{\bar\eta}\lambda_H\lambda_C,\nonumber\\
&&m_{\phi_\eta\phi_\eta} = m_\eta^2 + \frac{1}{4}\Big([(g_{YX}g_X + g_X^2)](v_d^2 - v_u^2)+2g_X^2(3v_\eta^2-v_{\bar\eta}^2)\Big) + \frac{\lambda_C^2}{2}(v_{\bar\eta}^2 + v_S^2),\nonumber\\
&&m_{\phi_d\phi_{\bar\eta}} = -\frac{1}{2}g_X(g_{YX} + g_X)v_dv_\eta -\frac{1}{2}v_uv_{\bar\eta}\lambda_H\lambda_C,\nonumber\\
&&m_{\phi_u\phi_{\bar\eta}} = \frac{1}{2}g_X(g_{YX} + g_X)v_dv_\eta -\frac{1}{2}v_dv_{\bar\eta}\lambda_H\lambda_C,\nonumber\\
&&m_{\phi_\eta\phi_{\bar\eta}} = ({\lambda}_C^2 - g_X^2)v_\eta v_{\bar\eta} + \frac{\lambda_C}{2}(2l_W - \lambda_Hv_dv_u) + \frac{v_S}{\sqrt{2}}(2M_S\lambda_C + T_{\lambda_C}) + \frac{v_S^2}{2}\lambda_C\kappa,\nonumber\\
&&m_{\phi_{\bar\eta}\phi_{\bar\eta}} = m_{\bar\eta}^2 + \frac{1}{4}\Big((g_{YX}g_X + g_X^2)(v_u^2 - v_d^2)+2g_X^2(3v_{\bar\eta}^2-v_\eta^2)\Big) + \frac{\lambda_C^2}{2}(v_\eta^2 + v_S^2),\nonumber\\
&&m_{\phi_d\phi_s} = \Big(\lambda_Hv_dv_S + \sqrt{2}v_d\mu - v_u(\kappa v_S + \sqrt{2}M_S)\Big)\lambda_H - \frac{1}{\sqrt{2}}v_uT_{\lambda_H},\nonumber\\
&&m_{\phi_u\phi_s} = \Big(\lambda_Hv_uv_S + \sqrt{2}v_u\mu - v_d(\kappa v_S + \sqrt{2}M_S)\Big)\lambda_H - \frac{1}{\sqrt{2}}v_dT_{\lambda_H},\nonumber\\
&&m_{\phi_\eta\phi_s} = \Big(\lambda_Cv_\eta v_S + v_{\bar\eta}(\kappa v_S + \sqrt{2}M_S)\Big)\lambda_C + \frac{1}{\sqrt{2}}v_{\bar\eta}T_{\lambda_C},\nonumber\\
&&m_{\phi_{\bar\eta}\phi_s} = \Big(\lambda_Cv_{\bar\eta} v_S + v_\eta(\kappa v_S + \sqrt{2}M_S)\Big)\lambda_C + \frac{1}{\sqrt{2}}v_\eta T_{\lambda_C},\nonumber\\
&&m_{\phi_s\phi_s} = m_S^2 + \Big(2l_W + 3v_S(\kappa v_S + 2\sqrt{2}M_S) + \lambda_Cv_\eta v_{\bar\eta} - \lambda_Hv_dv_u \Big)\kappa + 2B_S\nonumber\\&&\hspace{1.8cm} +\frac{1}{2}\lambda_C^2\xi^2 + \frac{1}{2}\lambda_H^2v^2 + 4M_S^2 + \sqrt{2}v_ST_\kappa.
\end{eqnarray}
This matrix is brought into diagonal form by $Z^H$
\begin{equation}
Z^HM_h^2Z^{H,\dagger} = M^{dia}_{2,h}.
\end{equation}

The trilinear couplings of CP-even Higgs bosons are essential for the triangle diagram contributions in $gg \to hh$ production. We present the explicit expression of the Higgs triple coupling $A_{HHH}$ for $H_i-H_j-H_k$ in the $U(1)_X$SSM as follows:
{\small\begin{eqnarray}
&&A_{HHH}=(\frac{1}{4}g_1^2+\frac{1}{4}g_{YX}^2+\frac{1}{4}g_2^2+\frac{1}{2}g_{YX}g_X+\frac{1}{4}g_X^2
-\lambda_H^2)[v_u\langle112\rangle+v_d\langle122\rangle]\nonumber\\&&
-(\frac{3}{4}g_1^2+\frac{3}{4}g_{YX}^2+\frac{3}{4}g_2^2+\frac{3}{2}g_{YX}g_X+\frac{3}{4}g_X^2)[v_u\langle111\rangle+v_d\langle222\rangle]
+\frac{1}{2}(g_{YX}g_X+g_X^2)\nonumber\\&&\times\Big[v_{\bar{\eta}}(\langle114\rangle+\langle224\rangle)-v_\eta(\langle113\rangle+\langle223\rangle)
+v_u(\langle233\rangle+\langle244\rangle)-v_d(\langle133\rangle+\langle144\rangle)
\Big]\nonumber\\&&-(v_S\lambda_H^2+\sqrt{2}\mu\lambda_H)(\langle115\rangle+\langle225\rangle)+
(\lambda_Hv_S\kappa+\sqrt{2}M_S\lambda_H+\frac{1}{\sqrt{2}}T_{\lambda_H})\langle125\rangle\nonumber\\&&
-(\lambda_Cv_S\kappa+\sqrt{2}M_S\lambda_C+\frac{1}{\sqrt{2}}T_{\lambda_C})\langle345\rangle
+\frac{1}{2}\lambda_H\lambda_C\Big[v_{\bar{\eta}}\langle123\rangle+v_{\eta}\langle124\rangle+v_u\langle134\rangle\nonumber\\&&+v_d\langle234\rangle\Big]
+(\lambda_Hv_u\kappa-v_d\lambda_H^2 )\langle155\rangle+(\lambda_Hv_d\kappa-v_u\lambda_H^2 )\langle255\rangle
-3g_X^2(v_\eta\langle333\rangle+v_{\bar{\eta}}\langle444\rangle)\nonumber\\&&+(g_X^2-\lambda_C^2)(v_\eta\langle344\rangle+v_{\bar{\eta}}\langle334\rangle)
-v_S\lambda_C^2(\langle335\rangle+\langle445\rangle)-(\lambda_C^2v_\eta+\lambda_Cv_{\bar{\eta}}\kappa)\langle355\rangle\nonumber\\&&
-(\lambda_C^2v_{\bar{\eta}}+\lambda_Cv_{\eta}\kappa)\langle455\rangle-(6v_S\kappa^2+6\sqrt{2}M_S\kappa+\sqrt{2}T_{\kappa})\langle555\rangle,\nonumber\\&&
\end{eqnarray}}
Here $\langle\alpha\alpha\alpha\rangle, \langle\alpha\alpha\beta\rangle, \langle\alpha\beta\gamma\rangle$ are the shorthand notations
{\small\begin{eqnarray}&&
 \langle\alpha\alpha\alpha\rangle=Z^H_{i\alpha}Z^H_{j\alpha}Z^H_{k\alpha},
~~~\langle\alpha\alpha\beta\rangle=Z^H_{i\alpha}Z^H_{j\alpha}Z^H_{k\beta}
+Z^H_{i\alpha}Z^H_{j\beta}Z^H_{k\alpha}+Z^H_{i\beta}Z^H_{j\alpha}Z^H_{k\alpha},(\alpha\neq\beta),\nonumber\\&&
\langle\alpha\beta\gamma\rangle=Z^H_{i\alpha}Z^H_{j\gamma}Z^H_{k\beta}+Z^H_{i\gamma}Z^H_{j\alpha}Z^H_{k\beta}
+Z^H_{i\alpha}Z^H_{j\beta}Z^H_{k\gamma}+Z^H_{i\gamma}Z^H_{j\beta}Z^H_{k\alpha}+Z^H_{i\beta}Z^H_{j\alpha}Z^H_{k\gamma}\nonumber\\&&\hspace{1.6cm}
+Z^H_{i\beta}Z^H_{j\gamma}Z^H_{k\alpha},~~~~(\alpha\neq\beta\neq\gamma).
\end{eqnarray}}

The coupling vertices between Higgs bosons and squark pairs are essential for calculating the one-loop amplitudes of the \(gg \to hh\) process. We first present the explicit expressions of the single-Higgs coupling vertices for both down-type and up-type squark sectors in Eqs.(\ref{a1},\ref{a2})
\begin{eqnarray}
&&A_{H \tilde{D} \tilde{D}}=\frac{1}{12}\Big\{ \sum_{a=1}^3Z_{j,a}^{D,*}Z_{k,a}^D \Big[(3 g^2_2+g_{YX} g_{X}+g^2_1+g^2_{YX}\!-\!12 Y^{2}_{d,a})v_d Z^H_{i1}-(3 g^2_2+g_{YX} g_{X}
\nonumber\\&&\hspace{1.2cm}+g^2_1+g^2_{YX})v_u Z^H_{i2}
+2 g_{YX} g_{X}(v_{\eta} Z^H_{i3}-v_{\bar{\eta}} Z^H_{i4}) \Big]
+\sum_{a=1}^3Z_{j,3+a}^{D,*}Z_{k,3+a}^D \Big[(2 g^2_1+2 g^2_{YX}
\nonumber\\&&\hspace{1.2cm} +3 g^2_{X}+5g_{YX} g_{X}-12 Y^{2}_{d,a})v_d Z^H_{i1}-(2 g^2_1+2g^2_{YX}+3 g^2_{X}+5g_{YX} g_{X})v_u Z^H_{i2}
\nonumber\\&&\hspace{1.2cm}+(4 g_{YX} g_{X}+6 g^2_{X})(v_{\eta} Z^H_{i3}-v_{\bar{\eta}} Z^H_{i4})\Big]+ \sum_{a=1}^3Z_{j,a}^{D,*}Z_{k,3+a}^D \Big[(-6\sqrt{2} T_{d,a} Z^H_{i1})+6(v_S {\lambda}^{*}_{H} Y_{d,a}
\nonumber\\&&\hspace{1.2cm}+\sqrt{2}{\mu}^{*}Y_{d,a})Z^H_{i2}+6 v_u {\lambda}^{*}_{H}Y_{d,a}Z^H_{i5}\Big]+\sum_{a=1}^3Z_{j,3+a}^{D,*}Z_{k,a}^D\Big[(-6\sqrt{2} T^{*}_{d,a} Z^H_{i1})+6(v_S {\lambda}_{H} Y^{*}_{d,a}
\nonumber\\&&\hspace{1.2cm}+\sqrt{2}{\mu}Y^{*}_{d,a})Z^H_{i2}+6 v_u {\lambda}_{H}Y^{*}_{d,a} Z^H_{i5}\Big] \Big\},\label{a1}\\
&&A_{H \tilde{U} \tilde{U}}=\frac{1}{12}\Big\{ \sum_{a=1}^3Z_{j,a}^{U,*}Z_{k,a}^U \Big[ (-3 g^2_2+g_{YX} g_{X}+g^2_1+g^2_{YX})v_d Z^H_{i1}-(-3 g^2_2+g_{YX} g_{X}+g^2_1
\nonumber\\&&\hspace{1.1cm} +g^2_{YX}-12 Y^{2}_{u,a})v_u Z^H_{i2}
+2 g_{YX} g_{X}(v_{\eta} Z^H_{i3}-v_{\bar{\eta}} Z^H_{i4}) \Big]+\sum_{a=1}^3Z_{j,3+a}^{U,*}Z_{k,3+a}^U  \Big[-(3 g^2_{X}
\nonumber\\&&\hspace{1.1cm}+4 g^2_1+4 g^2_{YX}+7 g_{YX} g_{X})v_d Z^H_{i1}+(3 g^2_{X}+4 g^2_1+4 g^2_{YX}+7 g_{YX} g_{X}-12 Y^{2}_{u,a})v_u Z^H_{i2}
\nonumber\\&&\hspace{1.1cm}-(8 g_{YX} g_{X}+6 g^2_{X})
(v_{\eta} Z^H_{i3}-v_{\bar{\eta}} Z^H_{i4}) \Big]
+ \sum_{a=1}^3Z_{j,a}^{U,*}Z_{k,3+a}^U \Big[6(v_S {\lambda}^{*}_{H} Y_{u,a}+\sqrt{2}{\mu}^{*}Y_{u,a})Z^H_{i1})
\nonumber\\&&\hspace{1.1cm} -6\sqrt{2} T_{u,a} Z^H_{i2}+6 v_d {\lambda}^{*}_{H}Y_{u,a}Z^H_{i5}\Big]+ \sum_{a=1}^3Z_{j,3+a}^{U,*}Z_{k,a}^U \Big[6(v_S {\lambda}_{H} Y^{*}_{u,a}+\sqrt{2}{\mu}Y^{*}_{u,a})Z^H_{i1})
\nonumber\\&&\hspace{1.1cm} -6\sqrt{2} T^{*}_{u,a} Z^H_{i2}+6 v_d {\lambda}_{H}Y^{*}_{u,a}Z^H_{i5}\Big] \Big\}.\label{a2}
\end{eqnarray}

We further give the explicit forms of the double-Higgs coupling vertices for both down-type and up-type squark sectors in Eqs.(\ref{a3},\ref{a4}).
\begin{eqnarray}
&&A_{HH \tilde{D} \tilde{D}}=\frac{1}{12}\Big\{ \sum_{a=1}^3Z_{k,a}^{D,*}Z_{l,a}^D \Big[ (3 g^2_2+g_{YX}g_{X}+g^2_1+g^2_{YX})(Z_{i1}^{H}Z_{j1}^{H}-Z_{i2}^{H}Z_{j2}^{H})
\nonumber\\&&\hspace{1.1cm}+2 g_{YX}g_{X}(Z_{i3}^{H}Z_{j3}^{H}-Z_{i4}^{H}Z_{j4}^{H})]+\sum_{a=1}^3Z_{k,3+a}^{D,*}Z_{l,3+a}^D[(3g^2_{X}+2g^2_1+2g^2_{YX}+5g_{YX}g_{X})
\nonumber\\&&\hspace{1.1cm}(Z_{i1}^{H}Z_{j1}^{H}-Z_{i2}^{H}Z_{j2}^{H})+2(3g^2_{X}+2g_{YX}g_{X})(Z_{i3}^{H}Z_{j3}^{H}-Z_{i4}^{H}Z_{j4}^{H})]
\nonumber\\&&\hspace{1.1cm}+6[-2\sum_{a=1}^3 Z_{k,3+a}^{D,*}Z_{l,3+a}^{D}Y^{2}_{d,a}Z_{i1}^{H}Z_{j1}^{H}-2\sum_{a=1}^3 Z_{k,a}^{D,*}Z_{l,a}^{D}Y^{2}_{d,a}Z_{i1}^{H}Z_{j1}^{H}
\nonumber\\&&\hspace{1.1cm}+(\lambda_H\sum_{a=1}^3Y_{d,a}^{*}Z_{k,3+a}^{D,*}Z_{l,a}^{D}+\lambda_H^{*}\sum_{a=1}^3Z_{k,a}^{D,*}Y_{d,a}Z_{l,3+a}^{D})Z_{i2}^{H}Z_{j5}^{H}-Z_{i5}^{H}Z_{j2}^{H}]\Big\},\label{a3}
\\
&&A_{HH \tilde{U} \tilde{U}}=\frac{1}{12} \Big\{ \sum_{a=1}^3Z_{k,a}^{U,*}Z_{l,a}^U \Big[ (-3 g^2_2+g_{YX}g_{X}+g^2_1+g^2_{YX})(Z_{i1}^{H}Z_{j1}^{H}-Z_{i2}^{H}Z_{j2}^{H})
\nonumber\\&&\hspace{1.1cm}+2 g_{YX}g_{X}(Z_{i3}^{H}Z_{j3}^{H}-Z_{i4}^{H}Z_{j4}^{H})]+\sum_{a=1}^3Z_{k,3+a}^{U,*}Z_{l,3+a}^U[(3g^2_{X}+4g^2_1+4g^2_{YX}+7g_{YX}g_{X})
\nonumber\\&&\hspace{1.1cm}(Z_{i2}^{H}Z_{j2}^{H}-Z_{i1}^{H}Z_{j1}^{H})-2(3g^2_{X}+4g_{YX}g_{X})(Z_{i3}^{H}Z_{j3}^{H}-Z_{i4}^{H}Z_{j4}^{H})]
\nonumber\\&&\hspace{1.1cm}+6[-2(\sum_{a=1}^3 Z_{k,3+a}^{U,*}Y^{2}_{u,a}Z_{l,3+a}^{U}+\sum_{a=1}^3 Z_{k,a}^{U,*}Y^{2}_{u,a}Z_{l,a}^{U})Z_{i2}^{H}Z_{j2}^{H}
\nonumber\\&&\hspace{1.1cm}+{\lambda}_H^{*}\sum_{a=1}^3Z_{k,a}^{U,*}Y_{u,a}Z_{l,3+a}^{U}(Z_{i1}^{H}Z_{j5}^{H}-Z_{i5}^{H}Z_{j1}^{H})]\Big\}.\label{a4}
\end{eqnarray}

\begin{acknowledgments}
This work is supported by National Natural Science Foundation of China (NNSFC)(No.12075074),
Natural Science Foundation of Hebei Province(A2020201002, A2023201040, A2022201022, A2022201017, A2023201041),
Natural Science Foundation of Hebei Education Department (QN2022173),
Post-graduate's Innovation Fund Project of Hebei University (HBU2024SS042),
This work is supported by the Project of the China Scholarship Council (CSC) No. 202408130113.
\end{acknowledgments}

\end{document}